\documentclass[global,twocolumn]{svjour}
\usepackage{latexsym}
\usepackage{graphics}
\usepackage{multirow} 
\usepackage{amsmath}
\journalname{Appl. Phys. B}
\begin{document}
\title{Generation of optical frequency combs via four-wave mixing processes for low- and medium-resolution astronomy}

\author{M. Zajnulina\inst{1} \and J. M. Chavez Boggio\inst{1} \and M. B\"{o}hm\inst{2} \and A. A. Rieznik\inst{3} \and T. Fremberg\inst{1} \and R. Haynes\inst{1} \and M. M. Roth\inst{1}
}                     

\institute{innoFSPEC-VKS, Leibniz Institute for Astrophysics Potsdam (AIP), An der Sternwarte 16, 14482 Potsdam, Germany \and innoFSPEC-InFaSe, University of Potsdam, Am M\"{u}hlenberg 3, 14476 Potsdam, Germany \and Instituto Tecnologico de Buenos Aires and CONICET, Buenos Aires, Argentina}
\date{Received: date / Revised version: date}

\maketitle
\begin{abstract}
We investigate the generation of optical frequency combs through a cascade of four-wave mixing processes in nonlinear fibres with optimised parameters.
The initial optical field consists of two continuous-wave lasers 
with frequency separation larger than $40~\text{GHz}$ ($312.7~\text{pm}$ at $1531~\text{nm}$). It propagates through three nonlinear 
fibres. The first fibre serves to pulse shape the initial sinusoidal-square pulse, while a strong pulse compression down to sub-$100~\text{fs}$ takes place in 
the second fibre which is an amplifying erbium-doped fibre. The last stage is a low-dispersion highly nonlinear fibre where the frequency comb bandwidth 
is increased and the line intensity is equalised. We model this system using the generalised nonlinear Schr\"{o}dinger equation and 
investigate it in terms of fibre lengths, fibre dispersion, laser frequency separation and input powers with the aim to minimise the frequency 
comb noise. With the support of the numerical results, a frequency comb is experimentally generated, first in the near infra-red and then it is 
frequency-doubled into the visible spectral range. Using a MUSE-type spectrograph, we evaluate the comb performance for astronomical wavelength 
calibration in terms of equidistancy of the comb lines and their stabi\-lity.
\end{abstract}

\section{Introduction}
\label{intro}
Optical frequency combs (OFCs) provide an array of phase-locked equidistant spectral lines with nearly equal intensity over a broad spectral range. Since their inception, they have triggered the development of a wide range of fields such as metrology for frequency synthesis \cite{cundiff}, for supercontinuum generation \cite{dudley,yang}, in the telecommunication for component testing, optical sampling, and ultra-high capacity transmission systems based on optical time-devision multiplexing \cite{pitois2,finot1,fortier,fatom2,mansouri,fatom}, or even for mimicking the physics of an event horizon \cite{webb}. 

One interesting application of OFCs is the calibration of astronomical spectrographs. Currently, wavelength ca\-libration of astronomical spectrographs uses the light of spectral emission lamps (Th/Ar, He, Ne, Hg, etc.) or absorption cells, for instance, iodine cells to map the dispersion 
function of the spectrograph \cite{griest}. These sources provide reliable and well characterised emission and absorption spectra, respectively, but have limitations in the spectral coverage. Moreover, because these lamps provide a line spacing and a line strengths that are irregular, the wavelength calibration 
accuracy is below optimal \cite{osterman,osterman2,ycas}. 

High-resolution applications like the search for extra-solar planets via the observation of the stellar radial velocities' Doppler shifts and the measurement of the cosmological fundamental constants require an accuracy of a few $\text{cm}/\text{s}$ in terms of radial velocity \cite{loeb,freedman,murphy1}.
The resolution of Th/Ar lamps is, however, limited to a few $\text{m}/\text{s}.$ 
Due to their properties, OFCs from mode-locked lasers were proposed as an ideal 
calibration source since they provide a much larger number of spectral lines at regions inaccessible for current lamps and with more equalised intensity \cite{osterman,phillips}. In has been demonstrated that broadband OFCs improved the accuracy by almost three orders of magnitude down to the $\text{cm}/\text{s}-$level. However, due to the tight spacing of their comb lines, mode-locked lasers have to be 
adapted using a set of stabilised Fabry-Perot cavities in order to increase their line spacing from hundreds of $\text{MHz}$ to $1-25~\text{GHz}.$ Frequency combs that were adapted using this technique have been successfully tested for high-resolution spectrographs $(R\geq 70000)$ in the visible and near infra-red (IR) \cite{osterman2,ycas,murphy1,phillips,murphy,braje,wilken,steinmetz,doerr,locurto}.
However, for low- and medium-resolution applications the filtering approach would require unfeasibly high-finesse stable Farby-Perot cavities to increase the spacing 
from hundreds of $\text{MHz}$ to hundreds of $\text{GHz}.$ 

Using monolithic microresonators, OFCs with a frequency line spacing between $100~\text{GHz}$ and $1~\text{THz}$ (suitable for the medium- and low-resolution 
range) have been recently demonstrated \cite{delhaye1,delhaye2}. However, due to the thermal effects, micro\-resonator-based combs cannot sustain the resonance condition for a long time and have to be regularly adjusted.

Another approach suitable for low- and medium-reso\-lution consists of generating a cascade of four-wave mixing (FWM) processes in optical fibres 
starting from two lasers. This allows OFCs to be generated with, in principle, arbitrary frequency spacing. This approach has been already 
extensively studied with the aim to generate ultra-short pulses at high repetition rates \cite{pitois2,finot1,fortier,fatom2,mansouri,fatom,chernikov}. But also some approaches specifically targeting the task of the OFC generation in highly nonlinear fibres were numerically and experimentally studied in the recent past \cite{tong_fwm,myslivets_fwm,yang_fwm}.  

We numerically investigate the four-wave mixing cascade approach with the particularity that it involves a long piece of an erbium-doped fibre with anomalous dispersion where strong pulse compression based on the higher-order soliton compression takes place \cite{boggio,zajnulina,zajnulina2}. We focus the analysis on how the quality of the compression and the pulse pedestal build-up depend on the input power, laser frequency separation, and group-velocity dispersion of the first fibre. We investigate how the intensity noise and the pulse coherence also depend on these parameters. Studies on the length optimisation of the first and second fibre stage allowing low-noise system performance are also carried out. Using a MUSE-type spectrograph, we experimentally demonstrate that the introduced approach is suitable for astronomical applications in the low- and medium-resolution range.
 
The paper is structured as follows: in Sec.~\ref{sec:setup}, we describe the approach for the generation of OFCs in fibres and, subsequently in Sec.~\ref{sec:model}, the according mathematical model based on the generalised nonlinear Schr\"{o}dinger equation (GNLS). We present our results on the fibre length optimisation in Sec.~\ref{sec:optimum_lenghts}. In Sec.~\ref{sec:figure_of_merit}, we show the results on the figure of merit and the pedestal content. The results of the noise evolution and coherence studies are shown in Sec.~\ref{sec:noise_evolution} and Sec.~\ref{sec:coherence}, respectively. In Sec.~\ref{sec:experimental_data}, we present the result on the experimental realisation of the proposed approach in the near IR and the visible spectral range. Finally, we draw our conclusions in Sec.~\ref{sec:conclusion}.

\section{Optical frequency comb approach and mathematical model}
\subsection{Four-wave-mixing based frequency comb}
\label{sec:setup}
\begin{figure}[htb]
\centering
\resizebox{0.48\textwidth}{!}{
\includegraphics{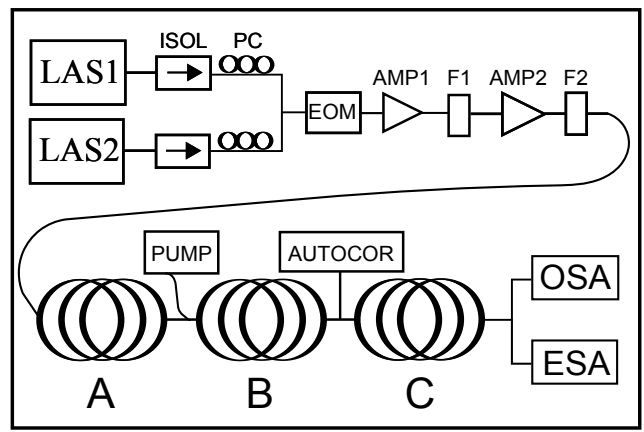}
}
\caption{Experimental setup for generation of OFCs in fibres. ISOL: optical isolator, PC: polarisation controller, EOM: electro-optical modulator, LAS1: fixed CW laser, LAS2: tunable CW laser, AMP1: Er-doped fibre amplifier 1, F1: optical bandpass filter 1, AMP2: Er-doped fibre amplifier 2, F2: optical bandpass filter 2, A: single-mode fibre, B: Er-doped fibre, C: highly nonlinear low-dispersion fibre, PUMP: pump laser for fibre B, AUTOCOR: optical autocorrelator, OSA: optical spectrum analyser, ESA: electrical spectrum analyser}
\label{fig:setup}
\end{figure}

Fig.~\ref{fig:setup} shows the experimental arrangement used to gene\-rate broadband optical frequency combs in the near IR spectral region. The starting optical field consists of two independent and free-running continuous-wave (CW) la\-sers. Both lasers have equal 
intensity and feature relative frequency stability of $10^{-8}$ over one-day time frame that is typical for state-of-the-art lasers. This stability is adequate for calibration of low- and medium-resolution astronomical 
spectrographs, no additional stabilising techniques like laser phase-locking are required. The first laser (LAS1) is fixed at the angular frequency $\omega_{1}$, while the second laser (LAS2) has a tuneable angular frequency $\omega_{2}$ so that the resulting modulated sine-wave has a central frequency $\omega_{c}=(\omega_{1}+\omega_{2})/2$.

\begin{figure}[htb]
\centering
\resizebox{0.35\textwidth}{!}{
\includegraphics{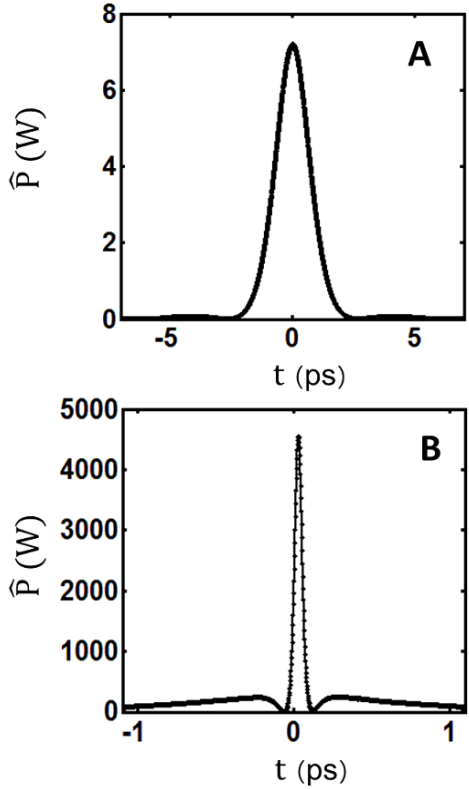}}
\caption{Optical pulse shapes after propagation in fibre A and B obtained by means of numerical simulations for laser frequency separation $LSF=80~\text{GHz}$ and initial power $P_{0}=2~\text{W}$}
\label{fig:time_dom}
\end{figure}

The evolution of a frequency comb in this system is governed by the following processes: as the two initial laser waves at $\omega_{1}$ and 
$\omega_{2}$ propagate through the fibre A, they interact through FWM and generate a cascade of new spectral components \cite{webb,cerqueira}. The new components are phase-correlated with the original laser lines, the frequency spacing between them coincides with the initial laser frequency separation $LFS=|\omega_{2}-\omega_{1}|/2\pi.$ In the time domain, this produce a moulding of the sinusoidal-square pulse: a train of well separated higher-order solitons with pulse widths of a few pico-seconds is gene\-rated \cite{mollenauer,haus}. These higher-oder solitons undergo further compression as they propagate through the amplifying fibre B \cite{colman,voronin,inoue}: sub-$100~\text{fs}$ pulses are generated (Fig.~\ref{fig:time_dom}) \cite{melo}. The last stage is a low-dispersion highly nonlinear fibre where the OFC gets broadened and the intensity of the comb lines fairly equalised.

\subsection{Generalised nonlinear Schr\"{o}dinger equation}
\label{sec:model}
We model the propagation of the bichromatic optical field using the ge\-ne\-ralised nonlinear Schr\"{o}dinger equation (GNLS) for a slowly varying amplitude $A=A(z,t)$ in the co-moving frame \cite{boggio,zajnulina,voronin,agrawal,agrawal2}:
\small
\begin{equation}
\frac{\partial A}{\partial z}=i\sum_{k=2}^{K}\frac{i^{k}}{k!}\beta_{k}\frac{\partial^{k}A}{\partial t^{k}}
+i\gamma\left(1+\frac{i}{\omega_{c}}\frac{\partial}{\partial t}\right)A\mathcal{R}
+g_{\text{Er}}A,
\label{eq:wave}
\end{equation}
\normalsize
where $\beta_{k}=\left(\frac{\partial^{k}\beta}{\partial\omega^{k}}\right)_{\omega=\omega_{c}}$ denotes the value of the dispersion order at the carrier angular frequency $\omega_{c}.$ The non\-li\-near parameter $\gamma$ is defined as $\gamma=\frac{\omega_{c}n_{2}}{cS}$ with $n_{2}$ being the nonlinear refractive index of silica, $S$ the effective mode area, and $c$ speed of light. The integral $\mathcal{R}=\int_{-\infty}^{\infty}R(t')|A(z,t-t')|^{2}dt'$ represents the response function of the nonlinear medium
\begin{equation}
R(t)=(1-f_{R})\delta(t)+f_{R}h_{R}(t),
\end{equation}
where the electronic contribution is assumed to be nearly instantaneous and the contribution set by vibration of silica molecules is expressed via $h_{R}(t).$ $f_{R}=0.245$ denotes the fraction of the delayed Raman response to the nonlinear polarisation. As for $h_{R}(t),$ it is defined as follows:
\begin{eqnarray}
h_{R}(t)=(1-f_{b})h_{a}(t)+f_{b}h_{b}(t),\\
h_{a}(t)=\frac{\tau^{2}_{1}+\tau^{2}_{2}}{\tau_{1}\tau_{2}^{2}}\exp\left(-\frac{t}{\tau_{2}}\right)\sin\left(\frac{t}{\tau_{1}}\right),\\
h_{b}(t)=\left(\frac{2\tau_{b}-t}{\tau_{b}^{2}}\right)\exp\left(-\frac{t}{\tau_{b}}\right)
\end{eqnarray}
with $\tau_{1}=12.2~\text{fs}$ and $\tau_{2}=32~\text{fs}$ being the characteristic times of the Raman response and $f_{b}=0.21$ representing the vibrational instability of silica with $\tau_{b}\approx 96~\text{fs}$ \cite{voronin,agrawal,agrawal2}. $g_{\text{Er}}$ in the last term on the right-hand side of Eq.~\ref{eq:wave} represents the normalised frequency-dependent Er-gain. Generally, $g_{\text{Er}}$ is a function of $z.$ Here, we use a gain profile that does not change with $z.$ This approach is justified by the fact that our numerical data are in a good agreement with the experimental ones. The Er-gain $g_{\text{Er}}$ is valid only for fibre B and is set to $g_{\text{Er}}=0$ for fibres A and C. 

\begin{figure}[htb]
\centering
\resizebox{0.38\textwidth}{!}{
\includegraphics{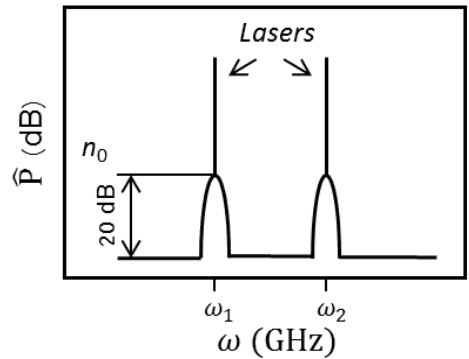}
}
\caption{Schematic representation of the initial condition}
\label{fig:noise_filter}
\end{figure}

The initial condition at $z=0$ for Eq.~\ref{eq:wave} reads as
\begin{equation}
A_{0}(t)=\sqrt{P_{0}}\sin(\omega_{c}t)+\sqrt{n_{0}(t)}\exp\left(i\phi_{\text{rand}}(t)\right),
\label{eq:IC}
\end{equation}  
where the first term describes the two-laser optical field with a peak power of $P_{0}$ and a central frequency $\omega_{c}=(\omega_{1}+\omega_{2})/2$ that coincides with the central wavelength of $\lambda_{c}=1531~\text{nm}.$ The second term in Eq.~\ref{eq:IC} describes the noise field and has the form of a randomly distributed floor with an amplitude varying between $0$ and $\sqrt{n_{0}}$ and a phase $\phi_{\text{rand}}$ randomly varying between $0$ and $2\pi.$ To mimic the experimental procedure in more detail, we convolve the noise floor with two filter functions having Gaussian shapes with a width of $30~\text{GHz}$ and a depth of $20~\text{dB}$ (see Fig.~\ref{fig:noise_filter}). The maximum of each Gaussian is positioned at the respective laser frequency line as shown in Fig.~\ref{fig:noise_filter}.   

The numerical solution of Eq.~\ref{eq:wave} having the initial optical field given by Eq.~\ref{eq:IC} is performed using the interaction picture method in combination with the local error method \cite{balac,cerqueira}. Low numerical error is obtained by choosing $2^{16}$ sample points in a temporal window of $256~\text{ps}.$ 

We consider up to the third order dispersion in our simulations, i.e. $K=3$ in Eq.~\ref{eq:wave}. Further, for the whole set of simulations, the following parameters for fibres A, B, and C are chosen:
$\gamma^{A}=2~\text{W}^{-1}\text{km}^{-1},$\\
$\beta_{2}^{B}=-14~\text{ps}^2/\text{km},$  $\gamma^{B}=2.5~\text{W}^{-1}\text{km}^{-1},$\\ $\beta_{2}^{C}=0.05~\text{ps}^{2}/\text{km},$ $\gamma^{B}=10~\text{W}^{-1}\text{km}^{-1}.$ \\The length of fibre C is set to  $L^{C}=1.27~\text{m}.$  These parameters represent material features of fibres that can be used in a real experiment.

\section{Optimum lengths of fibres A and B}
\label{sec:optimum_lenghts}
The aim of the propagation of the initial bichromatic field through fibres A and B is to generate maximally compressed optical pulses with a minimum level of intensity noise ($IN$). As the optical pulses propagate through fibres A and B, their intensity experiences periodical mo\-dulation over the pro\-pagation distance \cite{smyth}. This periodi\-city in the peak power occurs due to the formation and the subsequent propagation of higher-order solitons \cite{zajnulina2,kobtsev}.

\begin{figure}[htb]
\centering
\resizebox{0.36\textwidth}{!}{
\includegraphics{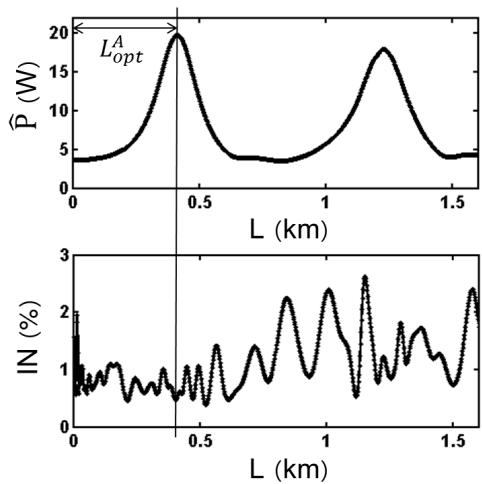}}
\caption{Peak power in $\text{W}$ (upper graph) and intensity noise in $\%$ (lower graph) vs. propagation distance in $\text{km}$ for fibre A}
\label{fig:1F_Why_IN}
\end{figure} 
We define the optimum length $L_{opt}$ of a fibre as the propagation distance between the beginning of the fibre and the first pulse intensity maximum. At the same time, the optimum length denotes the propagation distance point of the maximum optical-pulse compression and, thus, of the broadest possible spectrum \cite{li}. 

\begin{figure}[htb]
\centering
\resizebox{0.40\textwidth}{!}{
\includegraphics{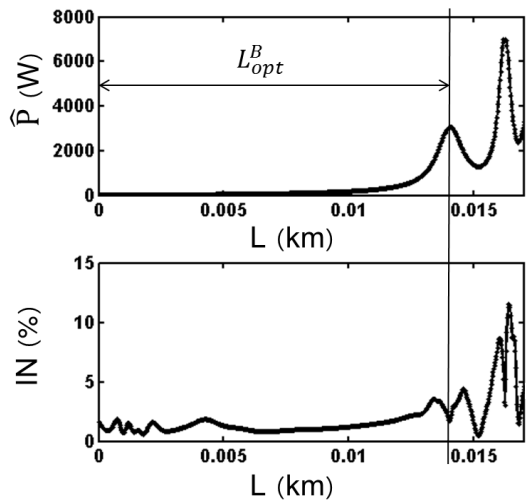}}
\caption{Peak power in $\text{W}$ (upper graph) and intensity noise in $\%$ (lower graph) vs. propagation distance in $\text{km}$ for fibre B}
\label{fig:2F_Why_IN}
\end{figure}

Rare-earth doped fibres are regarded as noisy environments and usually their lengths are kept as short as possible to avoid nonlinearities. Thus, the length optimisation studies provide us also with system parameters required to generate low intensity noise ($IN$) pulses and, so, low-noise OFCs. Fig.~\ref{fig:1F_Why_IN} and Fig.~\ref{fig:2F_Why_IN} show that the pulse $IN$ has a local minimum at optimum lengths of fibre A ($L_{opt}^{A}$) and B ($L_{opt}^{B}$). A more detailed discussion of intensity noise will be done in Sec.~\ref{sec:noise_evolution}.

To perform the optimisation studies, we assume the optical losses to be negligible, i.e. $\alpha=0~\text{dB/km}$ (see Eq.~\ref{eq:wave}).

\subsection{Optimum lengths of fibre A and B depending on the initial laser frequency separation}
\label{sec:L_opt_separation}
We consider three values of the initial laser frequency separation, i.e. $LFS=40~\text{GHz}$ ($312.7~\text{pm}$), $LFS=80~\text{GHz}$ ($625.5~\text{pm}$), and $LFS=160~\text{GHz}$ ($1.25~\text{nm}$ at $1531~\text{nm}$) that correspond to the medium and low resolution of $R=15000,$ $7500,$ and $3750$ at $1531~\text{nm}$ ta\-king into account that an optimum 
spacing between the comb lines is 3-4 times the spectrograph resolution (cf. \cite{murphy}). Having these values of $LFS,$ we look for optimum lengths of fibre A and B, i.e. $L_{opt}^{A}$ and $L_{opt}^{B},$ for different values of the input power $P_{0}$. For the studies, the group-velocity dispersion (GVD) parameter of fibre A is set to be $\beta^{A}_{2}=-15~\text{ps}^{2}/\text{km}.$ 
\begin{figure}[htb]
\centering
\resizebox{0.43\textwidth}{!}{
\includegraphics{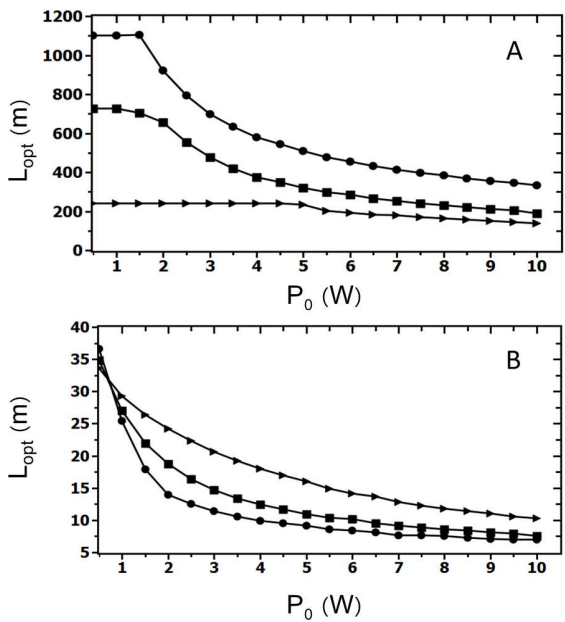}}
\caption{Optimum lengths of fibres A and B, $L_{opt}^{A}$ and $L_{opt}^{B},$ in $\text{m}$ vs. input power $P_{0}$ in $\text{W}$ for different values of the initial laser frequency separation $LFS:$ $LFS=40~\text{GHz}$ (circles), $LFS=80~\text{GHz}$ (rectangles), and $LFS=160~\text{GHz}$ (triangles)}
\label{fig:L_opt_separation}
\end{figure}

Fig.~\ref{fig:L_opt_separation} illustrates the dependence of optimum lengths on the input power $P_{0}.$ Generally, solitons with higher order numbers evolve on shorter lengths scales \cite{colman}. In our case, the soliton number can be calculated as
\begin{equation}
N^{A}=\sqrt{\frac{\gamma^{A}P_{0}}{(2\pi LFS)^{2}|\beta_{2}^{A}|}}
\label{eq:1F_N}
\end{equation}
for fibre A, or as
\begin{equation}
N^{B}=\sqrt{\frac{(T^{A}_{0})^{2}\hat{P}\gamma^{B}}{|\beta^{B}_{2}|}}
\label{eq:2F_N}
\end{equation}
for fibre B \cite{li}, where $\gamma^{A}$ denotes the nonlinear para\-meter of fibre A, $T^{A}_{0}\approx T^{A}_{\text{FWHM}}/1.763$ is the natural width of pulses after fibre A \cite{agrawal}, and $\hat{P}$ is the according  peak power. 
The dependence of $N$ on $\sqrt{P_{0}}$ or $\sqrt{\hat{P}}$ explains the decrease of $L_{opt}^{A}$ and $L_{opt}^{B}$ as the value of $P_{0}$ increases. 

For the case of fibre A, the decrease of optimum-length values is preceded by a plateau region where $L_{opt}^{A}$ is constant as a function of $P_{0}.$ In this region, $P_{0}$ is not sufficient to induce the nonlinearity that can effectively compress the initial sine wave into a train of solitons. The edge of the plateau ends denotes the the value of $P_{0}$ from which on the formation of solitons is fully supported. In terms of soliton order, $N^{A}<1$ within the plateau region and $N^{A}\geq 1$ for higher value of $P_{0}.$ 

According to Eq.~\ref{eq:1F_N}, the soliton order numbers in fibre A are inversely proportional to $LFS.$ For instance, we have $N^{A}=2.3$ for $LFS=40~\text{GHz},$ $N^{A}=1.6$ for $LFS=80~\text{GHz},$ and $N^{A}=1.15$ for $LFS=160~\text{GHz}$ at $P_{0}=5~\text{W}.$ Therefore, one would expect that $L_{opt}^{A}$ goes up with $LFS.$ As Fig.~\ref{fig:L_opt_separation} shows, this is not the case: $L_{opt}^{A}$ is inversely proportional to $LFS.$ We explain this phenomenon as follows: the level of complexity of the soliton's structure and the evolution behaviour grows with its order. For the initial sine-wave to be compressed into a train of solitons with higher order, it needs to propagate a longer distance so that the fibre nonlinearity can mould the pulses properly. In any case, more precise studies are needed to analyse the formation of higher-order solitons out of a sine-square wave.   

However, the soliton-oder scheme works perfectly for fibre B: $L_{opt}^{B}$ increases with $LFS.$ Fig.~\ref{fig:L_opt_separation} shows that the optimum lengths take the values $150~\text{m} <L_{opt}^{A}\leq 1100~\text{m}$ for fibre A, whereas for fibre B $7~\text{m} <L_{opt}^{A}< 35~\text{m}.$ Since in our case the optimum performance is shown for $LSF=80~\text{GHz},$ we will use this value for further studies.    
  
\subsection{Optimum lengths of fibre A and B depending on the group-velocity dispersion of fibre A}
\label{sec:L_opt_beta}
\begin{figure}[htb]
\centering
\resizebox{0.43\textwidth}{!}{
\includegraphics{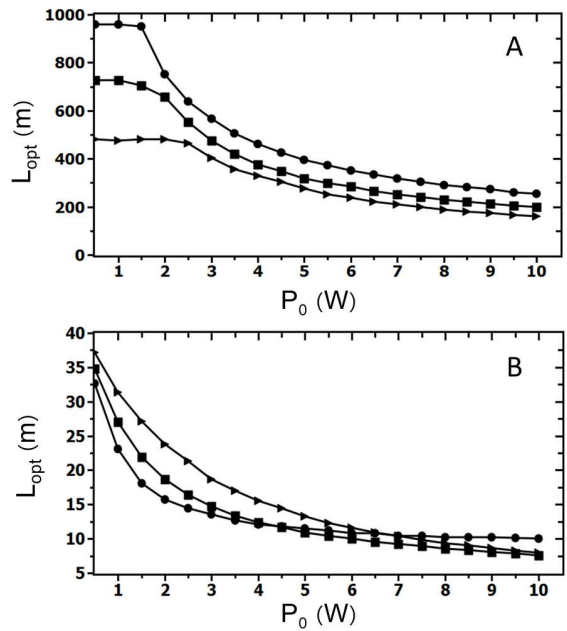}}
\caption{Optimum lengths of fibres A and B, $L_{opt}^{A}$ and $L_{opt}^{B},$ in $\text{m}$ vs. input power $P_{0}$ in $\text{W}$ for different values of the GVD parameter of fibre A: $\beta_{2}^{A}=-7.5~\text{ps}^{2}/\text{km}$ (circles), $\beta_{2}^{A}=-15~\text{ps}^{2}/\text{km}$ (rectangles), and $\beta_{2}^{A}=-30~\text{ps}^{2}/\text{km}$ (triangles)}
\label{fig:L_opt_beta}
\end{figure}

The dependency of the optimum fibre length as a function of three different values of the GVD parameter of fibre A is illustrated in Fig.~\ref{fig:L_opt_beta}. The three
dispersion values of fibre A are: $\beta_{2}^{A}=-7.5~\text{ps}^{2}/\text{km},$ $\beta_{2}^{A}=-15~\text{ps}^{2}/\text{km},$ and $\beta_{2}^{A}=-30~\text{ps}^{2}/\text{km}.$ These are standard values for single-mode fibres \cite{dudley,kobtsev2,kobtsev3}. The initial laser frequency separation is set to be $LFS=80~\text{GHz}.$ 

The optimum lengths for both, fibre A and B, decrease as the value of $P_{0}$ increases. Depending on the value of $P_{0},$ the optimum length of fibre A takes the values $180~\text{m}<L_{opt}^{A}<980~\text{m},$ for fibre B $7.5~\text{m}<L_{opt}^{B}<37.5~\text{m}.$ Also for different values of $\beta^{A}_{2},$ there plateaus of optimum length values for low input power. Precisely, the plateau region is $0.5~\text{W}\leq P_{0}<1.3~\text{W}$ for $\beta_{2}^{A}=-7.5~\text{ps}^{2}/\text{km},$ $0.5~\text{W}\leq P_{0}<1.5~\text{W}$ for $\beta_{2}^{A}=-15~\text{ps}^{2}/\text{km},$ and  $0.5~\text{W}\leq P_{0}<2.5~\text{W}$ for $\beta_{2}^{A}=-30~\text{ps}^{2}/\text{km}.$ In fibre A, the value of $L_{opt}^{A}$ increases as the absolute value of $\beta^{A}_{2}$ decreases, whereas it is the opposite dependency in fibre B. For further studies, we use, however, the value of $\beta_{2}^{A}=-15~\text{ps}^{2}/\text{km}.$

\section{Figure of merit and pedestal content}
\label{sec:figure_of_merit}
The higher-order soliton compression in an amplifying medium can be considered as an alternative technique to the compression in dispersion-decreasing 
fibres \cite{cao,chernikov2}. However, the compression of pico-second pulses suffers from the loss of the pulse energy into an undesired broad pedestal containing up to $70\%$ of the total pulse energy \cite{cao,li}. This has a reduction of the pulse peak power as a result leading to the degradation of the peak-power dependent FWM process. 

To describe the amount of energy that remains in the pulse and not in the pedestal, we introduce a figure of merit that is defined as:
\begin{equation}
FoM=\frac{\text{Pulse peak power}}{\text{Pulse average power}}.
\end{equation}  
Using the $FoM$, we address the following questions in this section:
\begin{itemize}
\item How does the $FoM$ of fibre B changes with the initial input power?
\item How does the $FoM$ of fibre B depends on the initial $LFS$ and $\beta_{2}^{A}?$
\item How the pedestal content depends on the the initial $LFS$ and $\beta_{2}^{A}?$ 
\end{itemize}

We define the pedestal content as a relative difference between the total energy of one single pulse and the energy of an approximating sech-profile with the same peak power and the FWHM as the pulse \cite{cao,li}:
\begin{equation}
PED=\frac{|E_{\text{total}}-E_{\text{sech}}|}{E_{\text{sech}}}\cdot 100\%.
\end{equation}
The sech-profile was chosen, because the pulses are molded into solitons in fibre A. The energy of a soliton with a sech-profile with peak power $\hat{P}$ and a FWHM is given by
\begin{equation}
E_{\text{sech}}=2\hat{P}\frac{\text{FWHM}}{1.763}.
\end{equation}

\subsection{Figure of merit and pedestal content of fibre B depending on the initial laser separation}
\label{sec:FoM_LSF}
To study the dependence of the figure of merit and the pedestal content in fibre B on the initial $LFS,$ we set again $LFS=40~\text{GHz},$ $LFS=80~\text{GHz},$ $LFS=160~\text{GHz}$ and $\beta_{2}^{A}=-15~\text{ps}^{2}/\text{km}.$   
\begin{figure}[htb]
\centering
\resizebox{0.44\textwidth}{!}{
\includegraphics{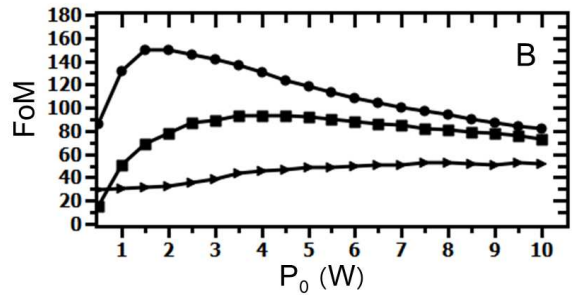}}
\caption{Figure of merit in fibre B for different values of the initial laser frequency separation $LFS:$ $LFS=40~\text{GHz}$ (circles), $LFS=80~\text{GHz}$ (rectangles), and $LFS=160~\text{GHz}$ (triangles)}
\label{fig:separation_FoM_2}
\end{figure}

Fig.~\ref{fig:separation_FoM_2} shows that the value of $FoM$ in fibre B is generally larger for smaller values of the initial $LFS.$ 
For $LSF=40~\text{GHz}$ and $LSF=80~\text{GHz},$ $FoM$ has a rapid increase for low input powers, reaches a maximum ($FoM=151$ at $P_{0}=1.5~\text{W}$ for $LSF=40~\text{GHz}$ and $FoM=93$ at $P_{0}=4~\text{W}$ for $LSF=80~\text{GHz}$) and starts to decrease as the value of $P_{0}$ increases further. 
A similar behaviour occurs for $LFS=160~\text{GHz}$ with a maximum lying beyond $P_{0}=10~\text{W}.$ 

\begin{figure}[htb]
\centering
\resizebox{0.44\textwidth}{!}{
\includegraphics{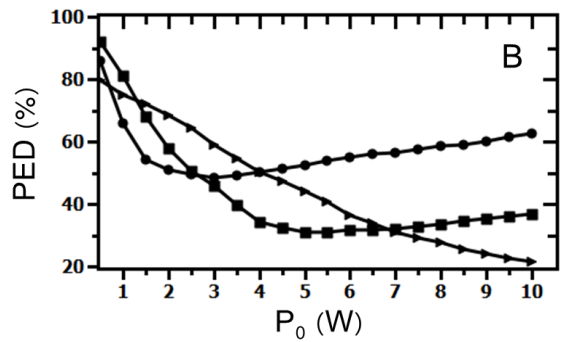}}
\caption{Pedestal energy content in fibre B in $\%$ for different values of the initial laser frequency separation $LFS:$ $LFS=40~\text{GHz}$ (circles), $LFS=80~\text{GHz}$ (rectangles), and $LFS=160~\text{GHz}$ (triangles)}
\label{fig:separation_pedestal}
\end{figure}

After decreasing of the pedestal content for low input powers, the value of $PED$ reaches a minimum ($PED=48.5\%$ at $P_{0}=3~\text{W}$ for $LFS=40~\text{GHz}$ and only $PED=30\%$ at $P_{0}=5~\text{W}$ for $LFS=80~\text{GHz}$) and then increases with $P_{0}$ again. The minima of $PED$ coincide with the soliton number $N>1.5$ of the pulses formed in fibre A. More precisely, $N=1.8$ for $LFS=40~\text{GHz}$ and $N=1.6$ for $LFS=80~\text{GHz}.$ Contrary to fundamental solitons with $N=1,$ any solitons with $N>1.5$ can be regarded as higher-oder solitons \cite{taylor}, the order will grow for higher input powers according to Eq.~\ref{eq:1F_N}. The increase of the pedestal content with $P_{0}$ presented in Fig.~\ref{fig:separation_pedestal} goes along with the increase of the soliton order numbers. This result is consistent with results published in Ref.~\cite{li}. In the considered input power region, $PED$ decreases continuously for $LFS=160~\text{GHz}$ reaching a value of only $PED=22\%$ for $P_{0}=10~\text{W}.$ The increase of $PED$ will occur for $P_{0}>10~\text{W}.$

A comparison of Fig.~\ref{fig:separation_FoM_2} and Fig.~\ref{fig:separation_pedestal} shows that the increase of $FoM$ for low input powers coincides with the decrease of $PED$ meaning that the most pulse energy gets effectively converted into the pulse peak power via the pulse compression. The increase of $PED$ causes the decrease of $FoM$ for higher values of $P_{0}.$

\subsection{Figure of merit and pedestal content of fibre B depending on the group-velocity dispersion of fibre A}
\label{sec:FoM_beta}
\begin{figure}[htb]
\centering
\resizebox{0.44\textwidth}{!}{
\includegraphics{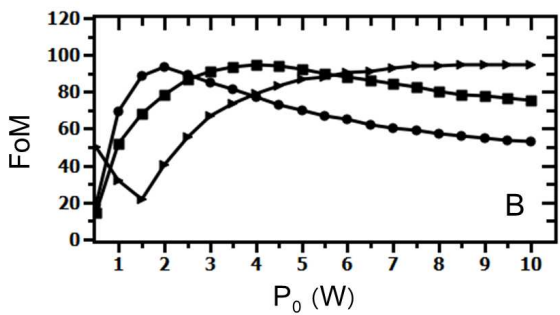}}
\caption{Figure of merit in fibre B for for different values of the GVD parameter of fibre A: $\beta_{2}^{A}=-7.5~\text{ps}^{2}/\text{km}$ (circles), $\beta_{2}^{A}=-15~\text{ps}^{2}/\text{km}$ (rectangles), and $\beta_{2}^{A}=-30~\text{ps}^{2}/\text{km}$ (triangles))}
\label{fig:beta2_FoM}
\end{figure}

Fig.~\ref{fig:beta2_FoM} shows that the maximum value of $FoM$ of fibre B does not depend on the GVD parameter chosen for fibre A. It shifts, however, to higher values of $P_{0}$ as the absolute value of $\beta_{2}^{A}$ increases ($FoM=93$ at $P_{0}=2~\text{W}$ for $\beta_{2}^{A}=-7.5~\text{ps}^{2}/\text{km}$ and $FoM=93$ at $P_{0}=4~\text{W}$ for $\beta_{2}^{A}=-15~\text{ps}^{2}/\text{km}$).
The decrease of $FoM$ after reaching a maximum is almost equally fast for $\beta_{2}^{A}=-7.5~\text{ps}^{2}/\text{km}$ $\beta_{2}^{A}=-15~\text{ps}^{2}/\text{km}.$ A similar behaviour will also occur for $\beta_{2}^{A}=-30~\text{ps}^{2}/\text{km}$ and higher values of $P_{0}.$

\begin{figure}[htb]
\centering
\resizebox{0.44\textwidth}{!}{
\includegraphics{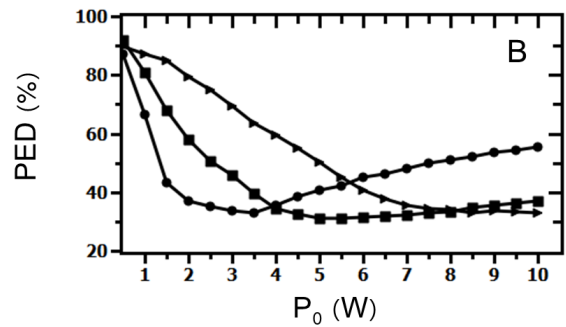}}
\caption{Pedestal energy content in fibre B in $\%$ for for different values of the GVD parameter of fibre A: $\beta_{2}^{A}=-7.5~\text{ps}^{2}/\text{km}$ (circles), $\beta_{2}^{A}=-15~\text{ps}^{2}/\text{km}$ (rectangles), and $\beta_{2}^{A}=-30~\text{ps}^{2}/\text{km}$ (triangles)}
\label{fig:beta2_pedestals_2}
\end{figure}

Fig.~\ref{fig:beta2_pedestals_2} shows that, again, the decrease of $FoM$ coincides with a build-up of the pedestal: after a minimum of only $PED=33\%$ for $\beta_{2}^{A}=-7.5~\text{ps}^{2}/\text{km}$ at $P_{0}=3.5~\text{W}$ $PED=30\%$ for $\beta_{2}^{A}=-15~\text{ps}^{2}/\text{km}$ at $P_{0}=5~\text{W},$ both curves start increasing. Thus, we have $PED=56.5\%$ for $\beta_{2}^{A}=-7.5~\text{ps}^{2}/\text{km}$ and $PED=38\%$ for $\beta_{2}^{A}=-15~\text{ps}^{2}/\text{km}$ at $P_{0}=10~\text{W}.$ The $PED-$minima coincide with soliton order of $N=1.9$ for $\beta_{2}^{A}=-7.5~\text{ps}^{2}/\text{km}$ and $N=1.6$ for $\beta_{2}^{A}=-15~\text{ps}^{2}/\text{km}.$ Again, the soliton order evolution causes the build-up of the pedestal. For $\beta_{2}^{A}=-30~\text{ps}^{2}/\text{km},$ the $PED-$curve decreases continuously as $P_{0}$ increases within the input power range we consider here, $PED=34\%$ at $P_{0}=10~\text{W}.$ 

Comparing the results obtained in Sec.~\ref{sec:FoM_LSF} and Sec.~\ref{sec:FoM_beta}, we see that the optimum system performance is obtained for $\beta_{2}^{A}=-15~\text{ps}^{2}/\text{km}$ and $LFS=80~\text{GHz}.$

\section{Intensity noise in Fibre A, B, and C}
\label{sec:noise_evolution}
The intensity noise ($IN$) coming from fibres A and B, can be strongly detrimental when the pulses propagate through fibre C. The high nonlinearity of this fibre increases the amount of the amplified noise of fibre B which leads to the reduction of the optical signal-to-noise ratio (OSRN) in the frequency domain. In this section, we investigate $IN$ in fibre B that comes from the amplification of any noise contributed from fibre A. In fibre A, the increase of intensity noise can be caused by modulational instability ~\cite{kobtsev}. 

The following questions are addressed here:
\begin{itemize}
\item How does the level of intensity noise in the amplifying fibre B, i.e. $IN^{B},$ depends on the initial $LFS$ and the value of the GVD of fibre A?
\item What importance has the initial $IN-$level for all three fibre stages?
\item How effective is the filtering technique consisting of two optical bandpass filters we proposed for the experiment?
\end{itemize}

We define the intensity noise $IN$ as the difference between the maximum peak power within a pulse train at the end of each fibre, i.e. $\text{max}(|\hat{A}|^{2}),$ and the according peak-power average, i.e. $\langle|\hat{A}|^{2}\rangle,$ in percentage terms:
\begin{equation}
IN=\frac{|\text{max}(|\hat{A}|^{2})-\langle|\hat{A}|^{2}\rangle|}{\langle|\hat{A}|^{2}\rangle}\cdot 100\%.
\end{equation}

Here, we consider three cases of the initial $IN-$power (Eq.~\ref{eq:IC}): the ideal case of $n_{0}=2P_{0}10^{-10}$ that conicides with $90~\text{dB}$ OSRN, $n_{0}=2P_{0}10^{-8}$ that corresponds to $70~\text{dB}$ OSRN, and $n_{0}=2P_{0}10^{-6}$ that corresponds to $50~\text{dB}$ OSRN. The first case is hardly realisable in a real experiment, while two latter ones are, on the contrary, realistic. We use optimised lengths of fibre A and B.

\subsection{Noise level in the amplifying stage depending on the initial laser frequency separation}
\label{sec:IN_separation}
To study of the intensity noise evolution in fibre B as a function of the initial $LFS,$ we chose the following va\-lues: $LFS=40~\text{GHz},$ $LFS=80~\text{GHz},$ $LFS=160~\text{GHz}.$ The initial intensity noise contribution is generated as a randomly distributed noise floor with the maximal power of $n_{0}=2P_{0}10^{-8}.$ The GVD parameter of fibre A is $\beta_{2}^{A}=-15~\text{ps}^{2}/\text{km}.$

\begin{figure}[htb]
\centering
\resizebox{0.32\textwidth}{!}{
\includegraphics{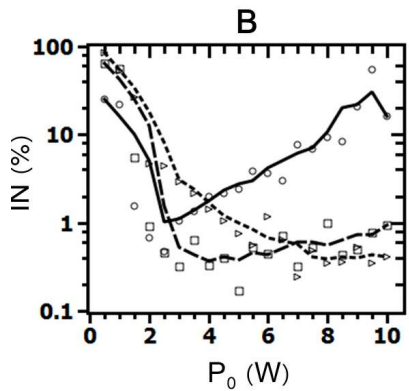}}
\caption{Intensity noise in fibre B, $IN^{B},$ in $\%$ vs. input power $P_{0}$ in W for different values of the initial laser frequency separation $LFS:$ $LFS=40~\text{GHz}$ (circles), $LFS=80~\text{GHz}$ (rectangles), and $LFS=160~\text{GHz}$ (triangles)}
\label{fig:IN_2F_Laser}
\end{figure}

Fig.~\ref{fig:IN_2F_Laser} shows that, for input powers for which fibre A has plateaus in its optimum lengths, the $IN^{B}-$level is very high (cf. Fig.~\ref{fig:L_opt_separation}). In this $P_{0}-$region, the optical pulses are not moulded into solitons yet when they propagate through fibre A (cf. Sec.~\ref{sec:optimum_lenghts}). Therefore, they lack the stability and robustness of real solitons to sustain the perturbation that is caused by the parameter change (GVD and nonlinearity) as they enter fibre B. As a result, the pulses break-up which yields a high level of $IN$ in fibre B. 

The resemblance of an optical pulse with a real soliton means its stability grow as the value of $P_{0}$ approaches the edge of the plateau region. So, the level of $IN^{B}$ decreases until it reaches a minimum at the plateau edge. Beyond the plateau region, the pulses are robust against the perturbation caused by the fibre parameter change since they are compressed to real solitons in fibre A. This has low intensity noise as a result: $IN<1\%$ for $LFS=80~\text{GHz}$ and $LFS=160~\text{GHz}.$ In Sec.~\ref{sec:L_opt_separation}, we showed that the soliton order is higher for smaller $LFS.$ Higher-order solitons are subjected to a break-up which leads to to the increase of intensity noise. This is why $IN$ increases up to ca. $10\%$ for $LFS=40~\text{GHz}.$ An optimal system performance is shown for $LSF=80~\text{GHz}.$

\subsection{Noise level in the amplyfying stage depenging on the group-velocity dispersion of fibre A}
\label{sec:IN_beta}
Having the maximal initial noise power of $n_{0}=2P_{0}10^{-8}$ generated as a floor and initial laser frequency separation of $LSF=80~\text{GHz},$ we now vary the GDV parameter of fibre A and choose the following values: $\beta_{2}^{A}=-7.5~\text{ps}^{2}/\text{km},$ $\beta_{2}^{A}=-15~\text{ps}^{2}/\text{km},$ $\beta_{2}^{A}=-30~\text{ps}^{2}/\text{km}.$ 
\begin{figure}[htb]
\centering
\resizebox{0.32\textwidth}{!}{
\includegraphics{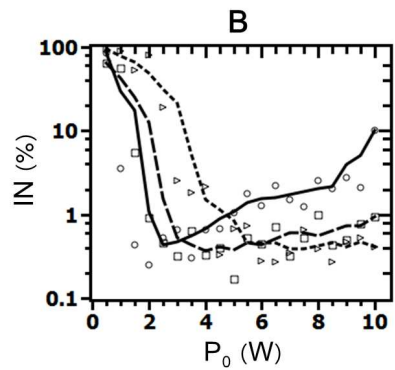}}
\caption{Intensity noise in fibre B, $IN^{B},$ in $\%$ vs. input power $P_{0}$ in $\text{W}$ for different values of the group-velocity dispersion of fibre A $\beta_{2}^{A}:$ $\beta_{2}^{A}=-7.5~\text{ps}^{2}/\text{km}$ (circles), $\beta_{2}^{A}=-15~\text{ps}^{2}/\text{km}$ (rectangles), and $\beta_{2}^{A}=-30~\text{ps}^{2}/\text{km}$ (triangles)}
\label{fig:IN_2F_beta}
\end{figure}

Fig.~\ref{fig:IN_2F_beta} shows that for input powers in the plateau region, the value of $IN^{B}$ is very high. Again, it occurs due to the instability and the resulting break-up of the optical pulses. For higher values of $P_{0},$ however, $IN^{B}$ remains below $1\%$ for $\beta_{2}^{A}=-15~\text{ps}^{2}/\text{km}$ and $\beta_{2}^{A}=-30~\text{ps}^{2}/\text{km}$ and increases up to $10\%$ for $\beta_{2}^{A}=-7.5~\text{ps}^{2}/\text{km}.$

As discussed in Sec.~\ref{sec:optimum_lenghts}, the soliton order grows as the absolute value of GVD of fibre A decreases. Higher-order solitons incline to the break-up for higher numbers of their order which has an increase of the intensity noise as a result. This is why we observe an increase of $IN^{B}$ up to $10\%$ for $\beta_{2}^{A}=-7.5~\text{ps}^{2}/\text{km}.$   

The best performance is shown for $\beta_{2}^{A}=-15~\text{ps}^{2}/\text{km},$ thus, we will use this value for further studies. 

\subsection{Intensity noise depending on the initial noise level. Effectiveness of the proposed filtering technique}
\label{sec:diff_noise}
For the study on the intensity noise of all fibre stages, i.e. $IN^{A},$ $IN^{B},$ and $IN^{C},$ we consider three cases of the initial $IN-$power generated as a randomly distributed floor (Eq.~\ref{eq:IC}): $n_{0}=2P_{0}10^{-10},$  $n_{0}=2P_{0}10^{-8}$ and $n_{0}=2P_{0}10^{-6}.$ The value of the frequency separation is chosen to be $LFS=80~\text{GHz}$ and the GVD parameter of fibre A is set to $\beta_{2}^{A}=-15~\text{ps}^{2}/\text{km}.$      

\begin{figure*}[htb]
\centering
\resizebox{1\textwidth}{!}{
\includegraphics{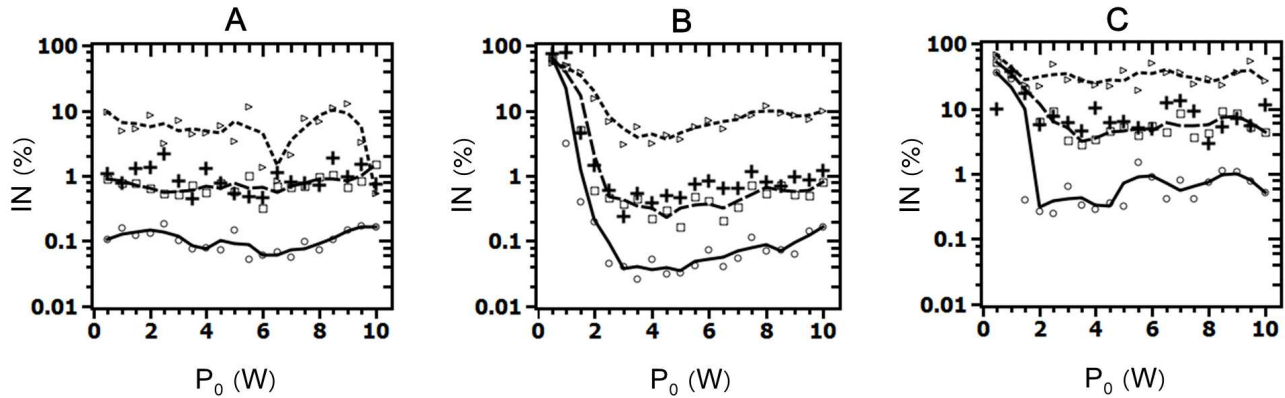}}
\caption{Intensity noise of fibres A, B, and C, $IN^{A},$ $IN^{B},$ and $IN^{C},$ in $\%$ vs. input power $P_{0}$ in $\text{W}$ for different values of the initial noise power: $n_{0}=2P_{0}10^{-10}$ (solid line), $n_{0}=2P_{0}10^{-8}$ (dashed line), and $n_{0}=2P_{0}10^{-6}$ (dotted line). The crosses present the intensity noise of the filtered signal.}
\label{fig:diff_noise}
\end{figure*}

Fig.~\ref{fig:diff_noise} shows that the whole system is sensitive to the value of the initial noise power. This dependence begins already in fibre A. Thus, $IN^{A}$ takes the following values: ca. $0.1\%$ for the ideal case of $n_{0}=2P_{0}10^{-10},$  ca. $1\%$ for $n_{0}=2P_{0}10^{-8},$ and ca. $10\%$ for $n_{0}=2P_{0}10^{-6}.$ 

In the fibre-A plateau region, the pulses not being real solitons yet and propagating through fibre B are extremely noisy for any values of $n_{0}$ due to their instability and the inclination to a break-up.
Although one would expect the intensity noise level to increase in the amp\-lifying fibre B, it actually gets slightly suppressed for $2.5~\text{W}\leq P_{0}<8~\text{W}.$ Apparently, fibre B has a stabi\-lising effect on the optical pulses in this input power region. For higher values of $P_{0},$ fibre B is not adding any addition noise, either. 

The nonlinearity of fibre C, however, adds a significant amount of $IN$ to the system, especially if the initial condition is highly noisy. So, we have $IN^{C}$ of $<1\%$ for $n_{0}=2P_{0}10^{-10},$ ca. $6\%$ for $n_{0}=2P_{0}10^{-8},$ and ca. $40\%$ for $n_{0}=2P_{0}10^{-6}$ for the values of $P_{0}-$region beyond the plateau region of fibre A. Thus, to keept the level of the intensity noise as low as possible it is advisable to choose a low-noise initial condition.  

Now we analyse the effectiveness of the proposed filtering technique. Two $20~\text{dB}-$filters with $30~\text{GHz}$ bandwidth was suggested to filter the noise coming from the amplifiers (AMP1 and AMP2 in Fig.~\ref{fig:setup}). The filters are modelled by two Gauss functions as described in Sec.~\ref{sec:model}. In our studies, the Gaussians filter the initial noise floor with $n_{0}=2P_{0}10^{-6}$ down to $n_{0}=2P_{0}10^{-8}$ (cf. Fig.~\ref{fig:noise_filter}). The according results are presented in Fig.~\ref{fig:diff_noise} as crosses. As one can see, the crosses lie close to the curves that present the $IN-$level for the situation when a noise floor with $n_{0}=2P_{0}10^{-8}$ is chosen as initial condition. To be precise, the $IN^{A}_{\text{filter}}$ is ca. $2\%,$ $IN^{B}_{\text{filter}}<1\%,$ and $IN^{C}_{\text{filter}}$ is less than $12\%$ for $P_{0}>2.5~\text{W}.$ That means that the proposed filtering technique is highly effective in the suppression of intensity noise and should be deployed in a real experiment. 

\section{Coherence in Fibre A, B, and C}
\label{sec:coherence}
The timing jitter of the optical pulses causes the broadening of the OFC lines. 
We study the impact of the timing jitter by means of the pulse coherence time $T_{c}$ that we define as the FWHM of the pulses that arise by a pairwise overlapping of pulse trains generated at two different times, $i$ and $i+1,$ and having, accordingly, different randomly generated initial $IN-$level. The overlap function is given by  
\begin{equation}
\tilde{g}(t)=\left\langle\frac{A^{*}_{i}(t)A_{i+1}(t)}{\sqrt{|A_{i}(t)|^{2}_{\text{max}}|A_{i+1}(t)|^{2}_{\text{max}}}}\right\rangle
\label{equ:coherence_2}
\end{equation}
where 
\begin{equation}
|A_{i}|^{2}_{max}=\text{max}(|A_{i}|^{2})
\end{equation}
is the maximum norm (cf.~\cite{agrawal}). For the calculation of $\tilde{g}(t),$ we use 10 different pulse trains, i. e. $i\in(1,...,10).$ A high level of pulse coherence corresponding to low timing jitter is presented when $T_{c}>T_{p}.$ Note, $T_{p}$ is the pulse FWHM.  

We consider the coherence time $T_{c}$ for three different va\-lues of the input power $P_{0}$ and initial noise $IN$ with $n_{0}=2P_{0}10^{-8}$ gene\-rated as a randomly distributed floor. Afterwards, these results will be compared with the case when the initial noise level with $n_{0}=2P_{0}10^{-6}$ is filtered down to $n_{0}=2P_{0}10^{-8}$ by means of Gaussian filters as described above. The initial frequency separation is chosen to be $LFS=80~\text{GHz}.$  
\begin{table}  
\begin{tabular}{ |l|l||l|l| }
  \hline
  \multicolumn{4}{|c|}{Fibre A} \\
  \hline
  $P_{0},$ $[W]$ & $IN$ & $T_{c},$ [ps] & $T_{p},$ [ps]\\
  \hline
  {$2.0$} & $\text{floor}$  & $6.16$ & $1.58$\\
  {} & $\text{filtered}$ & $6.39$ & {} \\ \hline

 {$5.5$} & $\text{floor}$  & $6.14$ & $0.67$\\
 {} & $\text{filtered}$ & $6.28$ & {} \\ \hline

 {$9.0$} & $\text{floor}$ & $6.15$ & $0.46$\\
 {} & $\text{filtered}$ &  $6.41$ & {} \\ \hline
\end{tabular}
\caption{Coherence time $T_{c}$ and FWHM of optical pulses $T_{p}$ in fibre A for a floor and filtered initial noise with $n_{0}=2P_{0}10^{-8}$}
\label{tab:coh_fib_A1}
\end{table}

As one notes from Tab.~\ref{tab:coh_fib_A1}, the pulse width $T_{p}$ decreases with the input power $P_{0}$ in fibre A due to the power-dependent compression process. Thus, we have $T_{p}=1.58~\text{ps}$ for $P_{0}=2.0~\text{W}$ and $T_{p}=0.46~\text{ps}$ for $P_{0}=9.0~\text{W}.$ However, for any values of $P_{0},$ the co\-herence time $T_{c}$ remains almost the same, it slightly varies around the average value of $\langle T_{c}\rangle =6.15~\text{ps}$ that is close to the natural pulse width of $T_{0}=6.4~\text{ps}$ in fibre A indicating high level of pulse coherence and very low timing jitter.
\begin{table} 
\begin{tabular}{ |l|l||l|l| }
  \hline
  \multicolumn{4}{|c|}{Fibre B} \\
  \hline
  $P_{0},$ $[W]$ & $IN$ & $T_{c},$ [ps] & $T_{p},$ [ps]\\
  \hline
   {$2.0$} & $\text{floor}$ & $1.56$ & $0.06$\\
   {} & $\text{filtered}$ & $1.61$ & {} \\ \hline

  {$5.5$} & $\text{floor}$ & $0.66$ & $0.08$ \\
  {}  & $\text{filtered}$ &  $0.67$ &  {} \\ \hline

  {$9.0$} & $\text{floor}$  & $0.46$ & $0.09$\\
  {} & $\text{filtered}$ & $0.47$ & {} \\ \hline
\end{tabular}
\caption{Coherence time $T_{c}$ and FWHM of optical pulses $T_{p}$ in fibre B for a floor and filtered initial noise with $n_{0}=2P_{0}10^{-8}$}
\label{tab:coh_fib_B1} 
\end{table}

In fibre B (Tab.~\ref{tab:coh_fib_B1}), the pulse widths $T_{p}$ slightly increase with the input power $P_{0}.$ This is the result of the decreasing compression effectiveness for the increasing input powers which we found out in further studies lying beyond the scope of this paper. So, we have $T_{c}=0.06~\text{ps}$ for $P_{0}=2.0~\text{W},$ $T_{c}=0.08~\text{ps}$ for $P_{0}=5.5~\text{W},$ and $T_{c}=0.09~\text{ps}$ for $P_{0}=9.0~\text{W}.$ 
Contrary to fibre A, the value of $T_{c}$ strongly depends  on the initial power: $T_{c}=1.56~\text{ps}$ for $P_{0}=2.0~\text{W},$  $T_{c}=0.66~\text{ps}$ for $P_{0}=5.5~\text{W}$, and finally  $T_{c}=0.46~\text{ps}$ for  $P_{0}=9.0~\text{W}.$ This occurs due to the fact that the pulse pedestal gets destroyed to a large extend as the input power increases. Nonetheless, the coherence $T_{c}$ is more than 5 times larger than the pulse width $T_{p}$ meaning still a good coherence performance with low timing jitter.   

\begin{table} 
\begin{tabular}{ |l|l||l|l| }
  \hline
  \multicolumn{4}{|c|}{Fibre C} \\
  \hline
 $P_{0},$ $[W]$ & $IN$ & $T_{c},$ [ps] & $T_{p},$ [ps]\\
  \hline
  {$2.0$}  & $\text{floor}$ & $0.07$ & $0.06$\\
  {} & $\text{filtered}$  & $0.08$ & {} \\ \hline

  {$5.5$}  & $\text{floor}$ & $0.08$ & $0.08$\\
  {} & $\text{filtered}$ & $0.09$ & {} \\ \hline
 
  {$9.0$} & $\text{floor}$  & $0.09$ & $0.09$\\
  {} & $\text{filtered}$  & $0.10$ & {} \\ \hline
\end{tabular}
\caption{Coherence time $T_{c}$ and FWHM of optical pulses $T_{p}$ in fibre C for a floor and filtered initial noise with $n_{0}=2P_{0}10^{-8}$}
\label{tab:coh_fib_C1} 
\end{table}

The optical pulses do not get compressed any further in fibre C (see Tab.~\ref{tab:coh_fib_C1}). However, the values of the coherence time $ T_{c}$ drop after the pulses propagated through fibre C and are only a bit higher than the pulse widths $T_{p}$: $T_{c}=0.07~\text{ps}$ for $P_{0}=2.0~\text{W},$ $T_{c}=0.08~\text{ps}$ for $P_{0}=5.5~\text{W},$ and $ T_{c}=0.09~\text{ps}$ for $P_{0}=9.0~\text{W}.$ The reason for low coherence time is the break-up of the pulse pedestal into pulses with irregular intensity and repetition due to the high fibre nonlinearity.

For the performed studies, the coherence time $T_{c}$ of the filtered signal lies slightly below the $T_{c}-$values of the unfiltered (floor) noise. This has only a negligible reduction of the coherent bandwidths. Thus, the proposed filtering technique proved to be effective once again. 

\section{Experimental data}
\label{sec:experimental_data}
Using the results from the numerical section where optimum fibre lengths, dispersion values, and input powers were found, we have setup an experimental arrangement to generate frequency combs for calibration of astronomical spectrographs. Fig.~\ref{fig:setup} shows the schematic of the experimental setup.

In the setup we used (cf. Fig.~\ref{fig:setup}), the EOM carves the initial wave that arises after the combination of both CW lasers into pulse trains with total extension of $20~\text{ns}.$ The first amplifier AMP1 provides an average power of $12~\text{mW}.$ The second amplifier AMP2 raises the average power to a value of $100~\text{mW}.$ The first filter F1 has a bandwidth of $100~\text{GHz},$ the bandwidth of the second filter F2 is $30~\text{GHz}.$ As the first stage (A) a conventional single-mode fibre with total length of $L^{A}= 350~\text{m}$ and the parameters $\beta_{2}^{A}=-21~\text{ps}^2/\text{km},$ $\gamma^{A}=2~\text{W}^{-1}\text{km}^{-1}$ was deployed. Instead of an Er-doped fibre (B), a double-clad Er/Yb-fibre with length of $L^{B}=17~\text{m}$ was used. This fibre got pumped with power of $3~\text{W}$ at $940~\text{nm}$. The fibre parameters are $\beta_{2}^{B}=-15~\text{ps}^2/\text{km},$ 
$\gamma^{B}=2.5~\text{W}^{-1}\text{km}^{-1}.$ Fibre C has the length of $L^{C}=3.5~\text{m}$ and the parameters $\beta_{2}^{C}=-0.5~\text{ps}^2/\text{km},$ 
$\gamma^{B}=10~\text{W}^{-1}\text{km}^{-1}$ at $1550~\text{nm}.$ The initial laser frequency separation was $LFS=200~\text{GHz}$ ($1.56~\text{nm}$ at $1531~\text{nm}$) which corresponded to the pulse repetition rate of $200~\text{GHz}$ in the time domain. 

\begin{figure}[htb]
\centering
\resizebox{0.50\textwidth}{!}{
\includegraphics{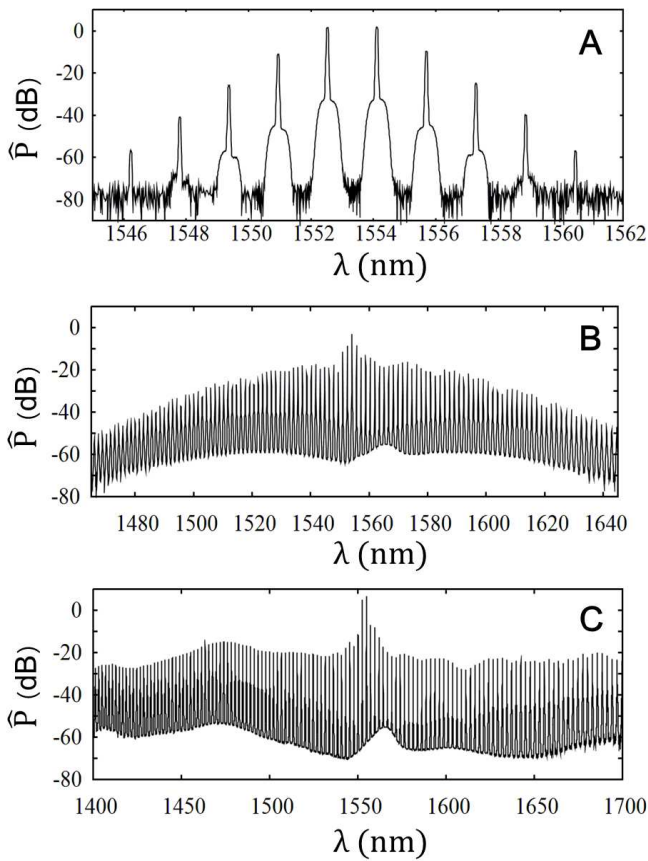}}
\caption{OFCs obtained after propagation through fibre stages A, B, and C with $LFS=200~\text{GHz}$}
\label{fig:comb_zusammen}
\end{figure}
Fig.~\ref{fig:comb_zusammen} shows typical spectra after fibre A, B, and C, respectively. The spectrum of fibre A ranges from $1546.2~\text{nm}$ to $1560.5~\text{nm},$ while the spectral bandwidth for fibre B is greatly extended from $1465~\text{nm}$ to $1645~\text{nm}.$ The line intensities in fibre A and B differ, however, in a few orders of magnitude. After propagation through fibre C, it is further broadened to the range between $1400~\text{nm}$ and $1700~\text{nm}$ and the line intensities are better equalised. Characterisation beyond $1700~\text{nm}$ was not possible due to limitations of the spectrometer used in the experiment. 

\begin{figure}[htb]
\centering
\resizebox{0.45\textwidth}{!}{
\includegraphics{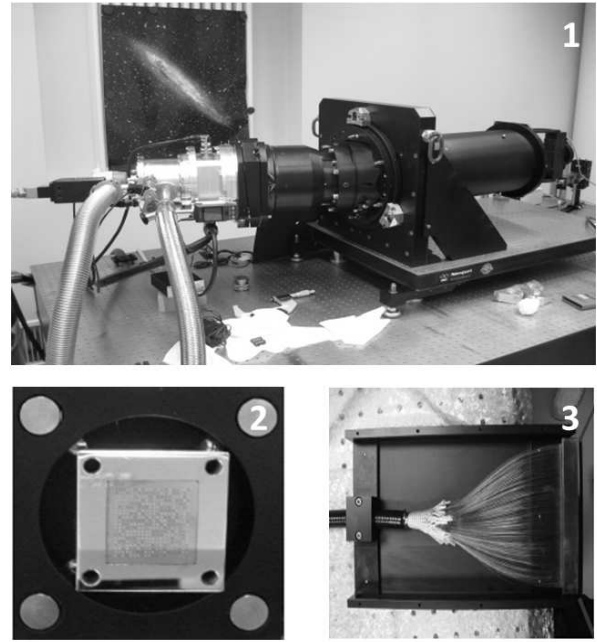}}
\caption{The MUSE-type spectrograph (1), the input (1) and the output (2) of the fibre bundle}
\label{fig:spec_in_put}
\end{figure}
To prove the effectiveness of the proposed system, we use a MUSE-type spectrograph (Fig.~\ref{fig:spec_in_put}.1). This spectrograph combines a broadband optical spectrograph with a new generation of multi-object deployable fibre bundles. It is a modified version of the Multi-Unit Spectroscopic 
Explorer (MUSE): instead of using image slicing mirrors, a $20\times 20$ fiber-fed input is used (Fig.~\ref{fig:spec_in_put}.2 and Fig.~\ref{fig:spec_in_put}.3). The MUSE instrument itself operates in the wavelength range between $465~\text{nm}$ to $930~\text{nm}$ with a $4096\times 4096$ CCD detector having $15~\mu\text{m}$ pixels. Its wavelength calibration is performed using the spectral lines from Ne and Hg lamps. The modified MUSE-type spectrograph we used exhibits the same features.

Thus, for the comb to be detectable by a MUSE-type spectrograph, we need to frequency-double the OFC obtained after fibre B into the visible spectral band. For that, an OFC centred at $1560~\text{nm}$ and spanning over $350~\text{nm}$ is focused into a BBO crystal with a thickness of $2~\text{mm}$ by means of a collimator and a focusing objective. 
\begin{figure}[htb]
\centering
\resizebox{0.49\textwidth}{!}{
\includegraphics{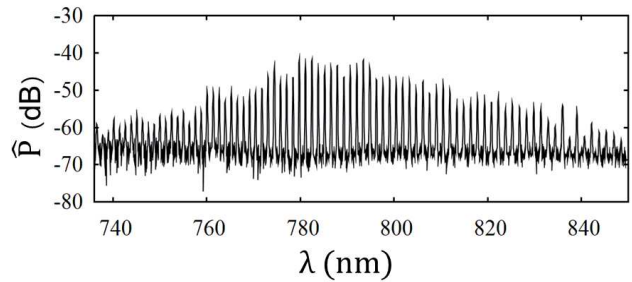}}
\caption{OFC obtained by means of the frequency-doubling of the output of fibre B}
\label{fig:comb_freq2_axis}
\end{figure}

Fig.~\ref{fig:comb_freq2_axis} shows the frequency-doubled spectrum obtained with $LFS=708~\text{GHz}$ ($5.54~\text{nm}$ at $1531~\text{nm}$). The spectrum extends from $736~\text{nm}$ to $850~\text{nm}$ and exhibits ca. 80 narrow equidistantly positioned lines. The lines have, however, different intensities which is caused by the frequency-doubling process. The frequency-doubling, however, has not imply a noticeable change of the coherence characteristics of the OFC. The best performance is in terms of the equality of line intensities is achievend in the spectral range between $780~\text{nm}$ and $800~\text{nm}.$

\begin{figure}[htb]
\centering
\resizebox{0.40\textwidth}{!}{
\includegraphics{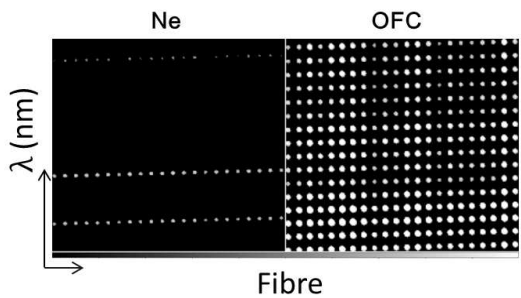}}
\caption{Comparison between calibration with a Ne lamp and an OFC in spectral region 1 of the MUSE-type spectrograph}
\label{fig:spectral_region1}
\end{figure}

\begin{figure}[htb]
\centering
\resizebox{0.40\textwidth}{!}{
\includegraphics{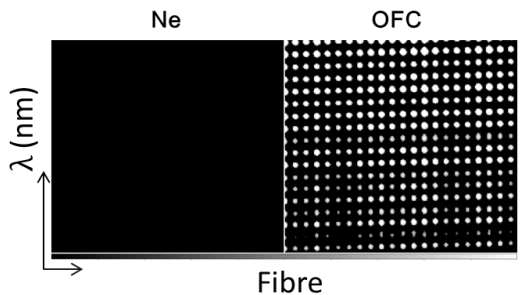}}
\caption{Comparison between calibration with a Ne lamp and an OFC in spectral region 2 of the MUSE-type spectrograph}
\label{fig:spectral_region2}
\end{figure}

A comparison between the calibration spectra of a Ne lamp and the frequency-doubled OFC was done using the MUSE-type spectrograph. The time exposure for both, the Ne and comb light, was $30~\text{s},$ while different exposures were taken with a few minutes of difference between them. Fig.~\ref{fig:spectral_region1} and Fig.~\ref{fig:spectral_region2} show the CCD images for two contiguous spectral regions (each one with $19.5~\text{nm}$ width) covering the range of $780-820~\text{nm}.$ Each comb line was sampled by 5 pixels. While the comb spectra exhibit bright and uniformly spaced peaks, the Ne light shows only three lines in the spectral region 1 and none in the other region.

In Sec.~\ref{sec:optimum_lenghts}, we drew our attention to the optimisation of the lengths of fibre A and B with the aim to achieve well-compressed optical pulses exhibiting minimal intensity noise. The lengths af stages A and B used for the experiment are close to the lengths obtained via numerial simulations. Thus, a good $IN-$performance was expected. However, the optical amplifiers add a large amount of $IN$ to the OCF. Nevertheless, the comb shows a good OSRN of more than $20~\text{dB}$ with the amount of optical power entering the spectrograph that is well above the detector's noise floor. 

\begin{figure}[htb]
\centering
\resizebox{0.50\textwidth}{!}{
\includegraphics{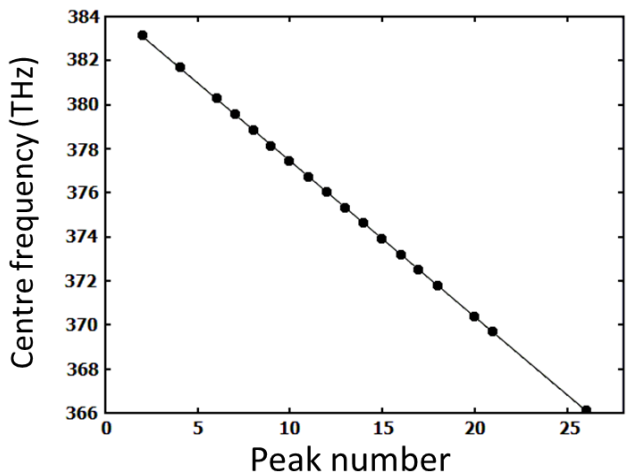}}
\caption{OFC line spacing for fibre no. 50 of the fibre bundle}
\label{fig:linespacing}
\end{figure}

To determine the line spacing between the comb lines, the detected light was reduced using a p3d software. Each line was independently fitted using a Gaussian function in order to have an accurate determination of the central wavelength and the line width. Fig.~\ref{fig:linespacing} shows the plot of the centre frequency as a function of the comb line.  

This was performed for all comb lines and for a representative number of the 400 fibres distributed over the field of view of the spectrograph. The results are summarised in Tab.~\ref{tab:line_spacing} for several fibres in the fibre bundle. As one can see, the line spacing changes from $708.5~\text{GHz}$ to $708.8~\text{GHz}$ among the
different fibres while the standard deviation for a single fibre is always $0.1~\text{GHz}.$ The main source of this deviation are the errors that arise during the fitting process with the help of Gaussian functions. Combs with this value of standard deviation are acceptable for astronomical application in the low- and medium resolution range. 
\begin{table}
\label{tab:line_spacing}  
\begin{tabular}{ |l|l|l| }
  \hline
  Fibre no. & Line spacing, [GHz] & Stand. dev., [GHz]\\ \hline
  $25$ & $708.6$ & $0.1$\\ \hline
  $45$ & $708.7$ & $0.1$\\ \hline
  $50$ & $708.5$ & $0.1$\\ \hline
  $51$ & $708.5$ & $0.1$\\ \hline
  $55$ & $708.5$ & $0.1$\\ \hline
  $100$& $708.7$ & $0.1$\\ \hline
  $150$& $708.8$ & $0.1$\\ \hline   
\end{tabular}
\caption{OFC line spacing for different fibres in the fibre bundle}
\end{table}

\section{Conclusion}
\label{sec:conclusion}
We investigated a fibre-based approach for generation of optical frequency combs via four-wave mixing in fibres starting from two CW lasers. This approach deploys an amplifying erbium-doped fibre stage. We performed numerical studies on the fibre length optimisation for different values of the input power $P_{0}$ ($0.5~\text{W}\leq P_{0}\leq 10~\text{W}),$ laser frequency separation $LSF,$ $LSF=40~\text{GHz}$ ($312.7~\text{pm}$), $80~\text{GHz}$ ($625.5~\text{pm}$), and $160~\text{GHz}$ ($1.25~\text{nm}$ at $1531~\text{nm}$), and the group-velocity dispersion parameter of the first fibre stage ($\beta_{2}^{A}=-7.5~\text{ps}^{2}/\text{km},$ $-15~\text{ps}^{2}/\text{km},$ $-30~\text{ps}^{2}/\text{km}$). Depending on the system parameters, the following fibre lengths were achieved via simulations: $150-1100~\text{m}$ for the first fibre stage and $7-37.5~\text{m}$ for the second (amplifying) stage. Since the simulations were performed neglecting the optical fibre losses, the real optimum length of the first fibre stage might be up to $50~\text{m}$ longer, for the second stage up to $10~\text{m}.$ 

The pulse compression in the amplifying fibre stage in our approach corresponds to the well-known higher-order soliton compression in dispersion-decreasing fibres. Using optimised fibre lengths, we showed that the undesired pulse pedestal content can be minimised to $30\%$ within the frame of our approach. Having introduced a figure of merit that describes the conversion of the pulse energy into the pulse peak power, we showed that the maximum of the figure of merit does not depend on the group-velocity dispersion parameter of the first fibre stage, but it is inversely proportional to the initial laser frequency separation. Accordingly, to achieve broad comb spectra, one should choose smaller laser frequency separation.

However, we also showed that smaller laser frequency separation leads to the higher intensity noise in the amp\-lifying stage. Our simulations showed that the intensity noise increases up to $10\%$ for the smallest value of the laser frequency separation chosen, i.e. for $40~\text{GHz}.$ For $80~\text{GHz}$ and $160~\text{GHz},$ it can be kept below $1\%.$ That means to achieve the best possible results, one needs to balance between the figure of merit and the noise performance. In our case, the optimum parameters were $LFS=80~\text{GHz}$ and $\beta_{2}^{A}=-15~\text{ps}^{2}/\text{km}.$     

Having chosen the optimum values, we studied the evolution of the intensity noise in all three fibre stages as a function of the initial intensity noise level. We showed that for the initial noise level that corresponds to $70~\text{dB}$ optical signal-to-noise ratio, the intensity noise in the first fibre is ca. $1\%$ for any values of the input power, is $<1\%$ in the amplifying fibre, and $<10\%$ in the third highly nonlinear fibre stage for input powers $>3~\text{W}.$ Moreover, we showed that the optical pulses exhibit high level of coherence in the first and second fibre stage and an acceptable level in the third one.    

We also showed that the proposed filtering technique that consists of two $20~\text{dB-}$filters with $30~\text{GHz}$ bandwidth is highly effective for the controlling of the intensity noise and the coherence properties of the system.   

Having used the numerical results, we generated a frequency comb to be used in an astronomical application. For that, we generated a frequency comb with laser frequency separation of $200~\text{GHz}$ ($1.56~\text{nm}$ at $1531~\text{nm}$) in all three fibre stages. To prove the equidistance of the comb lines, we deployed a MUSE-type spectrograph. For that, we frequency-doubled the comb (with frequency separation of $708~\text{GHz}$) ($5.54~\text{nm}$ at $1531~\text{nm}$) achieved after the second fibre into the visible spactral range. The comb that was detected by the MUSE-type spectrograph ranged between $780~\text{nm}$ and $820~\text{nm}.$ Having plotted the centroids of the comb lines, we realised that the standard deviation of the comb line spacing amounts to only $0.1~\text{GHz}$ ($0.8~\text{pm}$). In the course of further studies, we expect to generate a comb with a bandwidth of $150~\text{nm}$ at $800~\text{nm}$.      

To conclude, the approach we presented here is suitable for astronomical application in the low- and medium-resolution range in terms of noise and stability performance. A possible application taking advantage of our approach can be the 4MOST instrument addressing the research on the chemo-dynamical structure of the Milky Way, the cosmology with x-ray clusters of galaxies, and the Dark Energy \cite{dejong}.


\end{document}